\documentclass[a4paper]{jpconf}
\usepackage{graphicx}
\begin{document}
\title{The TREX-DM experiment at the Canfranc Underground Laboratory}

\author{J~Castel, S~Cebri\'an, T~Dafni, J~Gal\'an, IG~Irastorza, G~Luz\'on, C~Margalejo, H~Mirallas, A~Ortiz de Sol\'orzano, A~Peir\'o, E~Ruiz-Ch\'oliz}

\address{Centro de Astropart\'iculas y F\'isica de Altas Energ\'ias (CAPA), Universidad de Zaragoza, 50009 Zaragoza, Spain
\\ Laboratorio Subterr\'aneo de Canfranc, 22880 Canfranc Estaci\'on, Huesca, Spain}

\ead{scebrian@unizar.es}

\begin{abstract}
TREX-DM (TPC Rare Event eXperiment for Dark Matter) is intended to look for low mass WIMPs in the Canfranc Underground Laboratory (LSC) in Spain, using light elements (Ne, Ar) as target in a high pressure TPC equipped with Micromegas readouts.  Here, a description of the detector, the first results from commissioning data and the expected sensitivity from the developed background model are briefly presented.
\end{abstract}

\section{Detector and Commissioning data}
\begin{figure}[h]
\includegraphics[width=22pc]{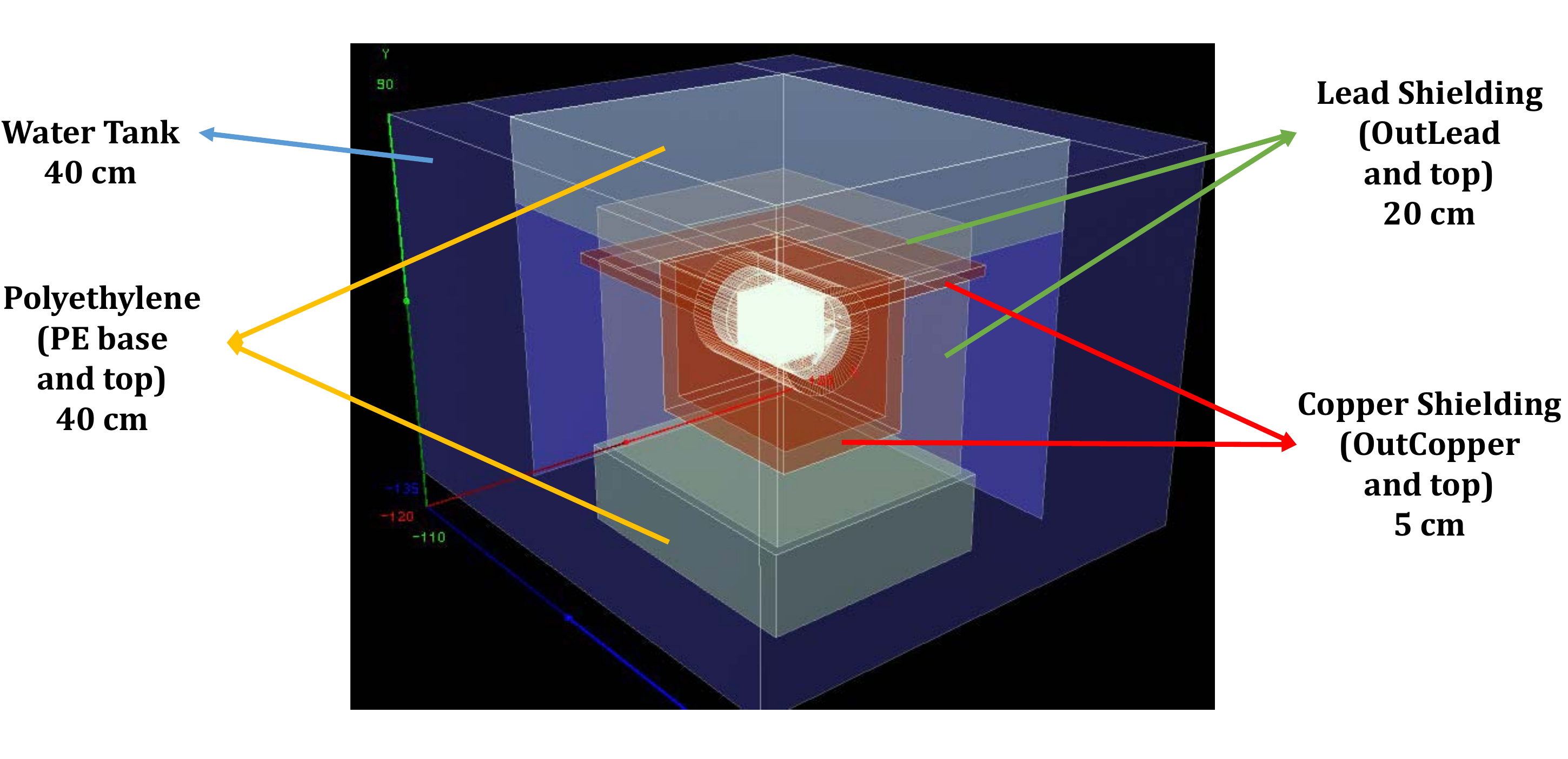}\hspace{2pc}%
\begin{minipage}[b]{14pc}\caption{\label{geometry}View of the TREX-DM set-up as simulated by GEANT4.}
\end{minipage}
\end{figure}

TREX-DM \cite{iguaz2016,irastorza2016} uses a gas TPC equipped with micromesh gas amplification structures (Micromegas), holding 24~l of gas at 10~bar (corresponding to 300~g of Ar, or alternatively, 160~g of Ne) inside a 6-cm-thick copper vessel with a central cathode defining two active volumes. The field cage is made of kapton and copper, covered by teflon. The two Micromegas readout planes, manufactured at CERN, have the largest surface (25$\times$25~cm$^{2}$) ever produced with the microbulk technique; they have low intrinsic radioactivity and allow to get topological information to discriminate signal from background. The detector is operated in the hall A of LSC (at 2450~m.w.e.) inside a shielding made of copper, lead and neutron moderator (see Fig.~\ref{geometry}); radon-free air is flushed inside the shielding. AGET-based electronics is being used, allowing to trigger from the low capacitance strips signals (256 in X, 256 in Y, with 1~mm pitch); then, an effective energy threshold at 100-400~eV$_{ee}$\footnote{Electron equivalent energy.} is expected.

\begin{figure}[]
\begin{minipage}{19pc}
\includegraphics[width=19pc]{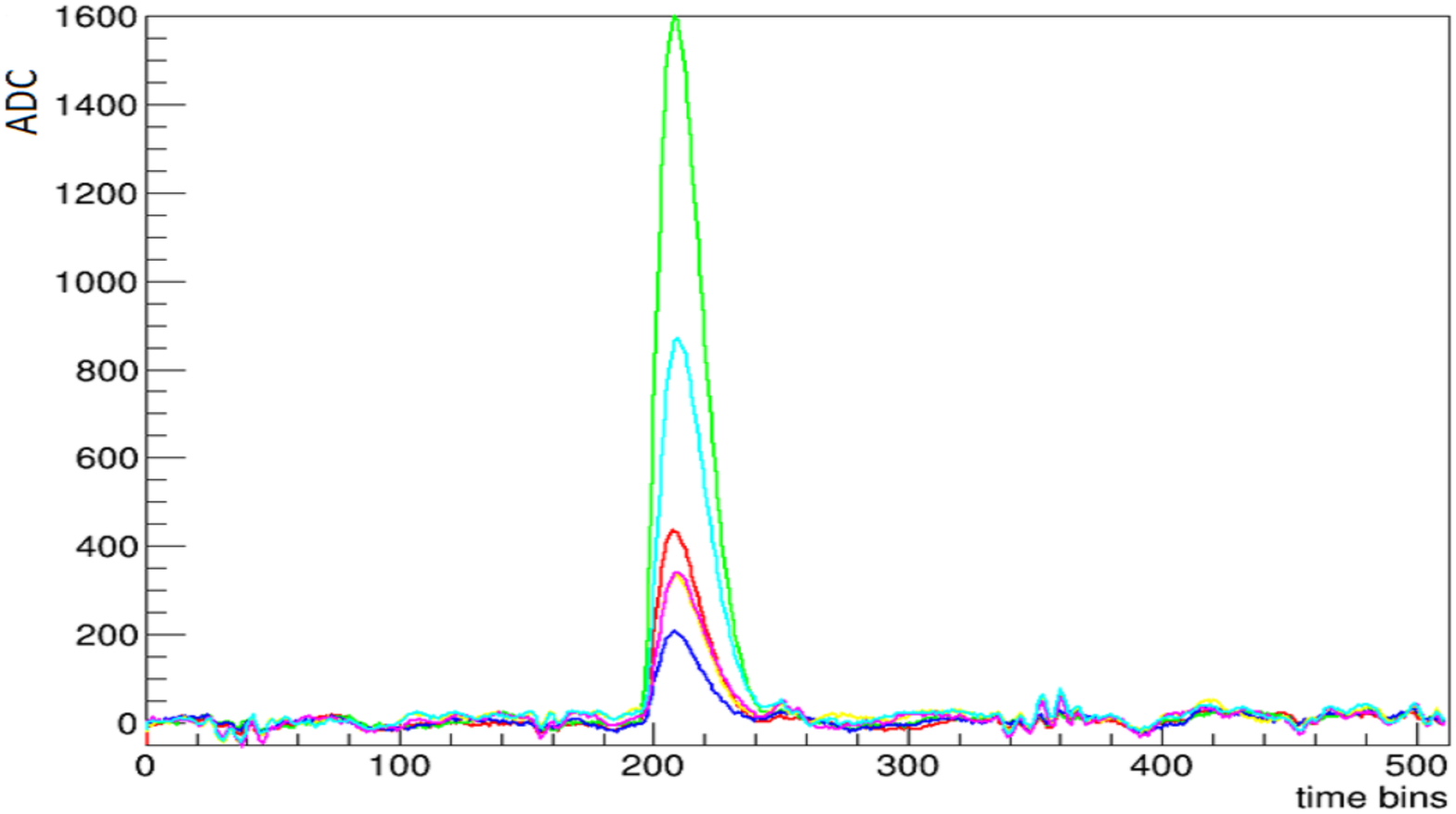}
\caption{\label{pulse}Example of strip pulses acquired for a 8~keV event.}
\end{minipage}\hspace{2pc}%
\begin{minipage}{18pc}
\includegraphics[width=18pc]{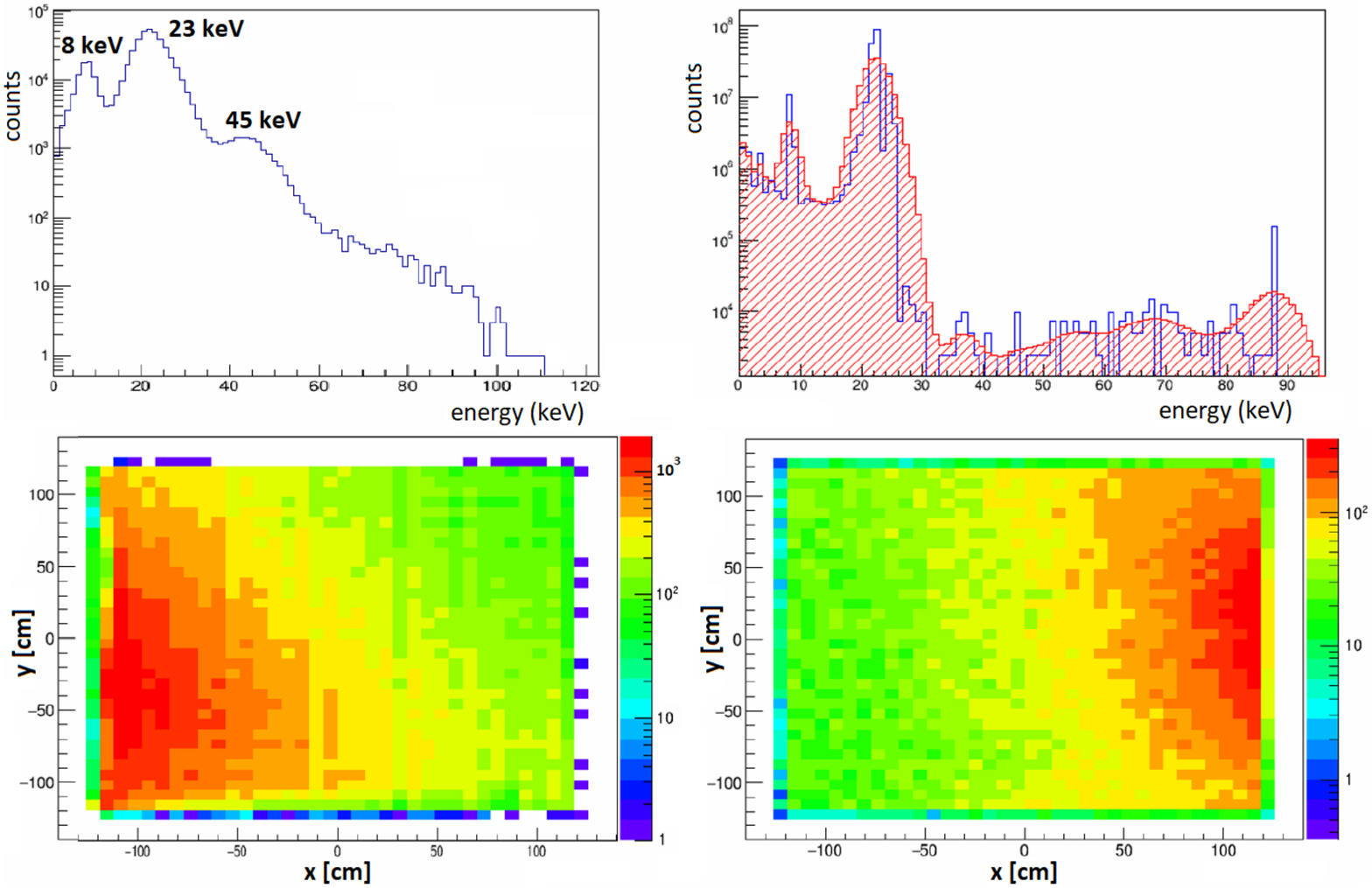}
\caption{\label{EnergySpectrum}Raw energy spectrum measured in a $^{109}$Cd calibration.}
\end{minipage}
\end{figure}

The detector was installed underground in Canfranc along 2018 and first tests with atmospheric Ar-1\%iC$_{4}$H$_{10}$ at 1.5~bar were made. In 2019, the gas was changed to the baseline mixture Ne-2\%iC$_{4}$H$_{10}$ working in recirculation, going progressively to higher pressures up to 8~bar in fall 2019. Despite some connectivity problems experienced, which were solved finally implementing face-to-face connections at the two readouts, the preliminary assessment of data seems positive. Following the calibration data and short background runs taken, rough calculations of the energy threshold confirm the expectations. Figure \ref{pulse} shows the acquired strip pulses for a 8~keV event from copper fluorescence and in Fig. \ref{EnergySpectrum} the raw energy spectrum measured in a $^{109}$Cd calibration using Ar is depicted; the structure at 45~keV is due to pile-up.

\section{Background model and Sensitivity}
\begin{figure}[h]
\includegraphics[width=22pc]{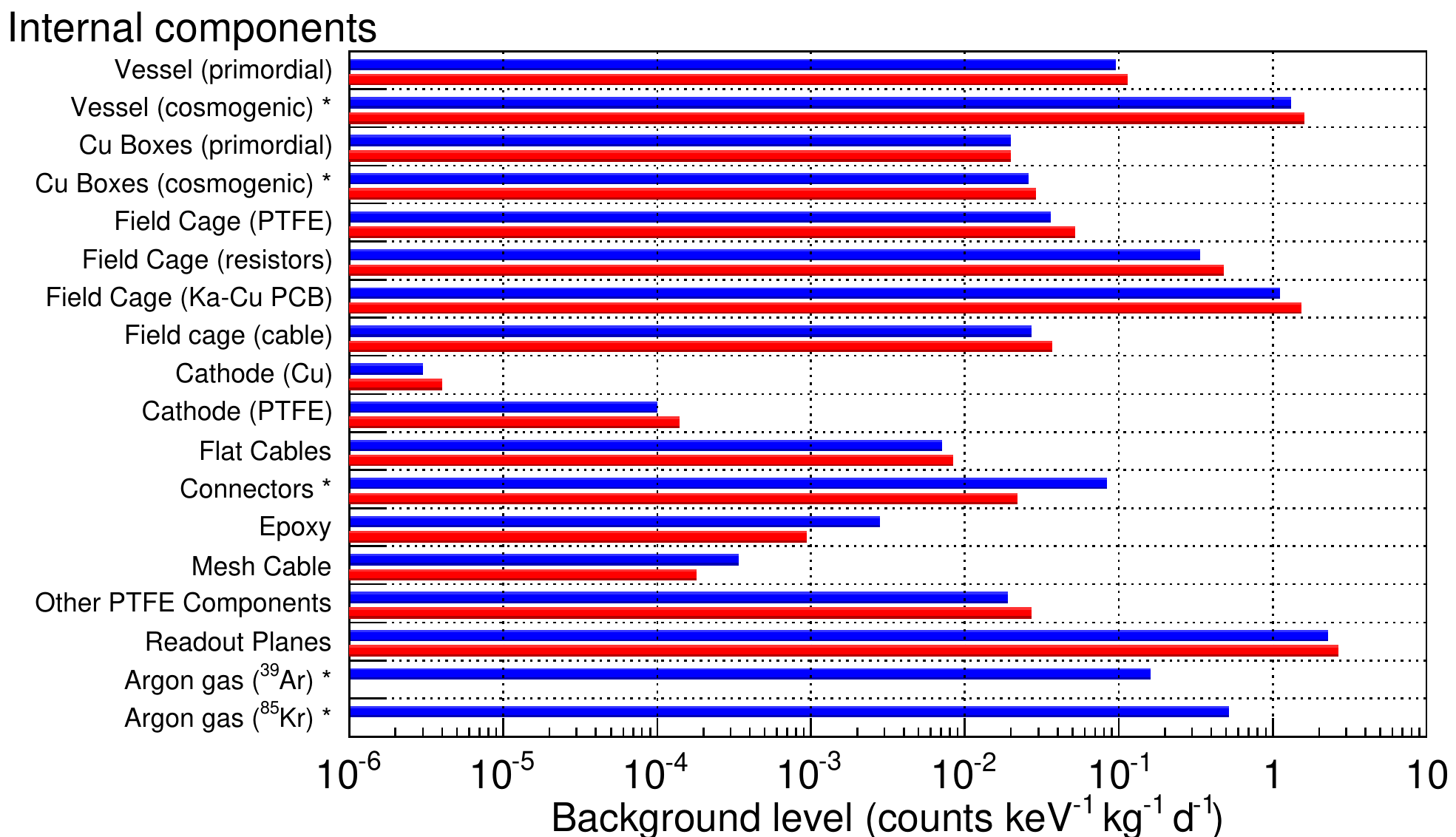}\hspace{2pc}%
\begin{minipage}[b]{14pc}\caption{\label{model}Background rates expected from internal components in the region 0.2-7~keV$_{ee}$ using Ar (blue bars) or Ne (red bars) mixtures in TREX-DM, as estimated in \cite{castel2019}.}
\end{minipage}
\end{figure}

An exhaustive radioassay program has been performed over the last years to construct the detector with radiopurity specifications, mainly based on germanium gamma spectrometry at LSC and complementary measurements from GDMS, ICPMS and the BiPo-3 detector. The background model developed by Monte Carlo simulation for the present set-up points to levels at 1-10~keV$^{-1}$ kg$^{-1}$ d$^{-1}$ in the region of interest \cite{castel2019}. It is based on Geant4 (for Physics processes) and a custom-made code called REST (for electron generation, diffusion effects, charge amplification and signal generation). Material radioactivity and environmental backgrounds in Canfranc (gamma-rays, neutrons and muons) have been considered. In the region from 0.2 to 7~keV$_{ee}$, the simulated contribution from internal components (at or inside the vessel) is $<$6.1(6.6)~keV$^{-1}$ kg$^{-1}$ d$^{-1}$ for Ar(Ne), while that of external components is $<$0.43(0.56)~keV$^{-1}$ kg$^{-1}$ d$^{-1}$ for Ar(Ne). As shown in Fig.~\ref{model}, the two main quantified contributions (the activated copper vessel and $^{40}$K in Micromegas) should be suppressed to get a background at the level of 1~keV$^{-1}$ kg$^{-1}$ d$^{-1}$. A roadmap to further decrease it down to 0.1~keV$^{-1}$ kg$^{-1}$ d$^{-1}$ is underway.


\begin{figure}[h]
\includegraphics[width=22pc]{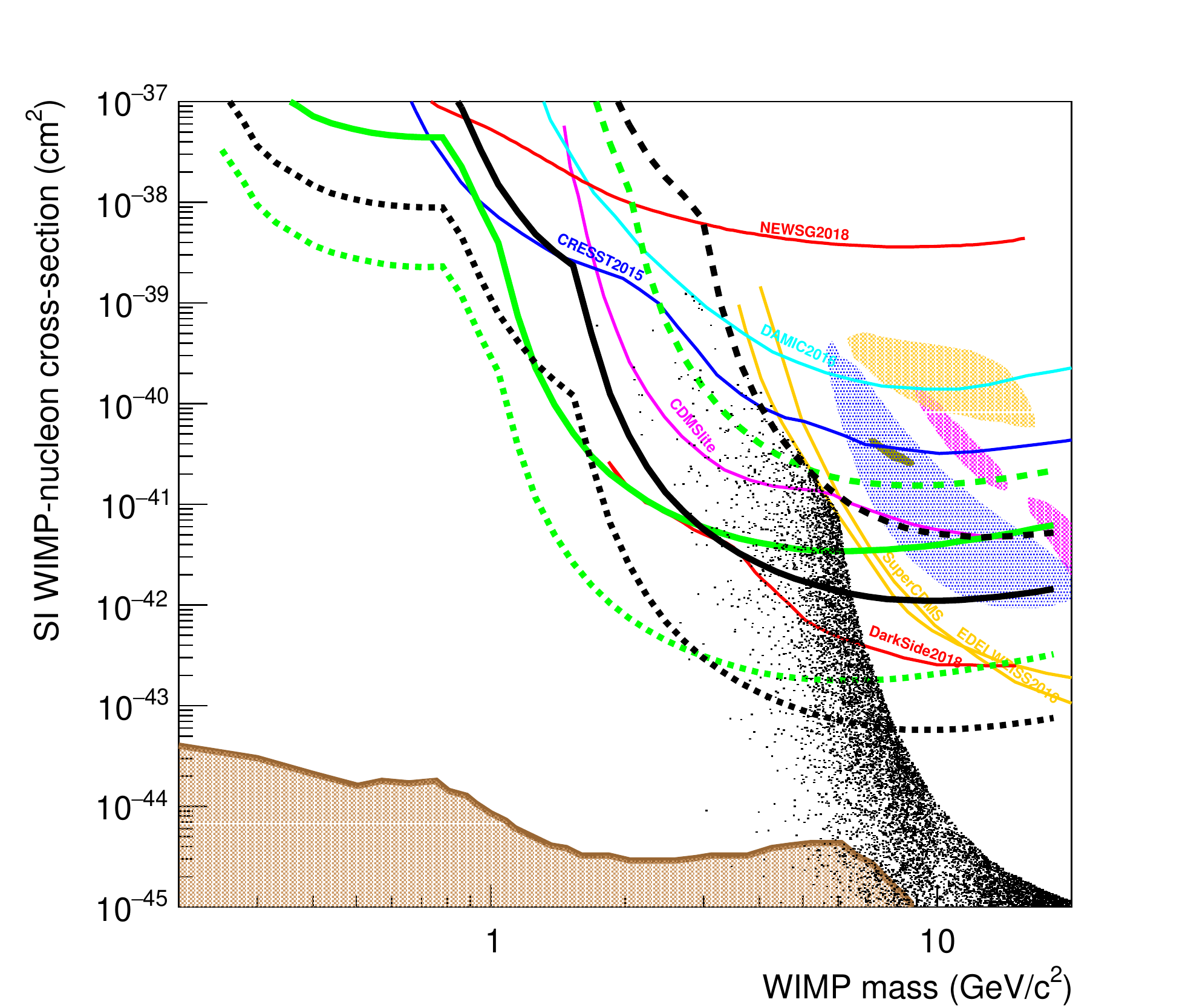}\hspace{2pc}%
\begin{minipage}[b]{14pc}\caption{\label{exclusion} 90\% C.L. sensitivity of TREX-DM under three different experimental conditions (see text) for Ar (black lines) and Ne (green lines) \cite{castel2019}; exclusion limits from different experiments are also shown and the cloud of points corresponds to viable light WIMPs compatible with collider constraints.}
\end{minipage}
\end{figure}

Prospects for exclusion plots have been evaluated for Spin-Independent interaction and standard halo model for three different experimental scenarios at a pressure of 10~bar, assuming different background levels (10, 1 and 0.1~keV$^{-1}$ kg$^{-1}$ d$^{-1}$), energy thresholds (0.4, 0.1 and 0.1~keV$_{ee}$) and exposures (0.3, 0.3 and 10~kg y) \cite{castel2019} and are shown in Fig.~\ref{exclusion}. Energy conversion between electronic and nuclear recoils is based on the commonly used parametrization in terms of the atom and mass numbers \cite{castel2019}. The best assumed conditions would provide a competitive sensitivity in the direct detection of low mass WIMPs. Operation with Ne first and underground Ar afterwards is foreseen as soon as the commissioning phase is finished.

\ack This work has been financially supported by the European Commission under the European Research Council T-REX Starting Grant ref. ERC-2009-StG-240054 of the IDEAS program of the 7th EU Framework Program and by the Spanish Ministry of Economy and Competitiveness (MINECO) under Grants FPA2013-41085-P and FPA2016-76978-C3-1-P.

\end{document}